\documentclass[journal]{IEEEtran}

\usepackage[caption=false,font=footnotesize]{subfig}
\usepackage{xcolor}

\usepackage{hyperref}
\usepackage[linesnumbered, ruled]{algorithm2e}

\SetCommentSty{mycommfont}

\usepackage{amsmath}
\usepackage{amssymb}
\usepackage{commath}
\usepackage{multirow}

\usepackage{tabularx}

\ifCLASSINFOpdf
  \usepackage[pdftex]{graphicx}

  \DeclareGraphicsExtensions{.pdf,.png}
\else
\fi

\begin{document}
\bstctlcite{IEEEexample:BSTcontrol}

\title{DENIS-SDN: Software-Defined Network Slicing Solution for Dense and Ultra-Dense IoT Networks}

\author{Tryfon~Theodorou,~\IEEEmembership{Member,~IEEE,} and~Lefteris~Mamatas,~\IEEEmembership{Member,~IEEE}
\thanks{The authors are with the Department of Applied Informatics, University of Macedonia, 156 Egnatia Street, GR-546 36 Thessaloniki, Greece (e-mails: \{theodorou, emamatas\}@uom.edu.gr).}}

\maketitle

\begin{abstract}
Traditional Wireless Sensor Networks protocols used in Internet of Things Networks (IoTNs) today face challenges in high- and ultra-density network topology conditions.
New networking paradigms like Software-Defined Networks (SDN) have emerged as an up-and-coming approach to address IoT application requirements through implementing global protocol strategies and network programmability. 
This paper proposes a divide-and-conquer solution that aims to improve the PDR in ultra-dense IoT (UDIoT) network environments using network slicing.
As such, we develop and evaluate \emph{DENIS-SDN}, an open-source SDN solution for UDIoT Network environments consisting of a modular SDN controller and an OpenFlow-like data-plane protocol. 
\emph{DENIS-SDN} utilizes our Network Density Control mechanism based on operational specification requirements, which address the challenges UDIoT network deployments pose, including interference, congestion, resource management, control, and quality of service (QoS) performance issues. 
To achieve this, it divides dense IoT networks into either logically sliced sub-networks separating nodes using routing rules or physically sliced sub-networks separating nodes into different radio channels. 
We provide evaluation results over realistic scenarios demonstrating improved PDR performance up to $4.8\%$ for logically and up to $11.6\%$ for physically sliced network scenarios.
\end{abstract}


\IEEEpeerreviewmaketitle

\section{Introduction} \label{seq:introduction} 
The Internet of Things (IoT) has opened up new possibilities for enhancing everyday life by enabling applications with demanding communication requirements, including massive IoT deployments. The use of IoT has transformed various domains, including smart cities, healthcare, transportation, and industrial automation, by connecting devices and collecting data for analysis and decision-making. These applications often have unique communication demands, such as stringent latency, reliability, and scalability requirements. Nowadays, a sub-domain of IoT networks that is gaining momentum, the Ultra-Dense IoT networks (UDIoT), are characterized by a large number of IoT devices deployed in a small geographic area, resulting in a high density of devices per unit area~\cite{yu2016ultra}. These networks can have hundreds to thousands of devices per square kilometer. They are commonly found in domains such as smart cities, industrial automation, smart grids, and transportation systems. Ultra-dense IoT networks pose unique challenges, including radio interference, congestion, scalability, energy efficiency, Quality of Service (QoS) provisioning, resource management and control, as well as standardization and interoperability~\cite{kamel2016ultra},~\cite{teng2018resource}. Addressing these challenges is crucial for UDIoT networks' efficient and effective operation, unlocking their full potential for supporting innovative IoT applications in various domains. However, state-of-the-art routing protocols for IoT, like Low-Power and Lossy Networks (RPL), are experiencing difficulties in efficiently handling these demands due to limitations in their design, such as the overhead of maintaining routing tables, lack of flexibility in routing decisions, and challenges in adapting to changing network conditions.

To overcome such limitations, researchers have proposed integrating the Software-Defined Networking (SDN) paradigm with IoT networks~\cite{sood2016SDNWSN-survey}, \cite{bera2017SDNWSN-survey}. SDN is a networking architecture that decouples the control plane from the data plane, allowing for centralized control and management of network resources~\cite{kreutz2014SDN-survey}. By leveraging the programmability and flexibility of SDN, IoT networks can benefit from more informed and customized routing control. For UDIoT networks' operation in particular, SDN can provide significant support by enabling dynamic and efficient control over the network resources, improving flexibility and manageability. In detail, the SDN paradigm can support UDIoT in the following areas: 
\begin{itemize}
    \item \emph{Network resource orchestration}, utilizing centralized control and management of network resources, enabling dynamic allocation and orchestration of resources in real-time.
    \item \emph{Dynamic traffic control}, allowing for intelligent and adaptive routing decisions based on real-time network conditions, optimizing path selection, and avoiding congestion.
    \item \emph{Network slicing}, allowing the creation of virtualized and isolated network segments tailored to the specific requirements of different IoT applications or services. 
    \item \emph{Centralized management and control}, providing a global view of the network state and enabling coordinated network-wide decisions. 
    \item \emph{Programmability and flexibility}, enabling rapid customization and adaptation of network behavior to changing requirements with quick adjustments and optimizations, ensuring efficient network operation.
    \item \emph{Network monitoring and analytics}, allowing real-time monitoring of network performance, traffic patterns, and resource utilization enabling proactive network management and early detection of anomalies or issues.
\end{itemize}
Nevertheless, SDN introduces additional complexities not encountered in conventional IoT networks, such as the amplified quantity of control packets that can detrimentally impact radio communication quality. To address these challenges, our SDN framework for IoT networks, \emph{CORAL-SDN}, originally introduced in demo paper~\cite{theodorou2017coralsdn}, proposes innovative centralized topology discovery mechanisms~\cite{theodorou2017topology}. Whereas our recent SDN solutions for IoT networks, \emph{VERO-SDN}~\cite{theodorou2020VERO-SDN} and \emph{SD-MIoT}~\cite{theodorou2020SD-MIoT}, demonstrate reduced control overhead while ensuring efficient routing and forwarding decisions with minimal end-to-end communication delays for both static and mobile IoT environments. However, to our knowledge, none of the existing solutions consider scenarios of IoT networks with dense or ultra-dense deployments.

In this context, we propose \emph{DENIS-SDN}, an open-source framework~\cite{DENIS-SDN2023github} aiming to address the challenges that dense and ultra-dense network deployments bring to IoT, including interference, congestion, resource management and control, and QoS performance issues.
In line with the SDN paradigm, the \emph{DENIS-SDN} framework dissociates control complexity from the network protocol and offloads it to a modular SDN Controller deployed in the surrounding fixed infrastructure. \emph{DENIS-SDN} implements programmable network slicing using its \emph{Network Density Control (NDC)} mechanism, providing dedicated and customized resources, routing policies, and QoS provisioning for different applications, ensuring efficient and reliable network operation in terms of PDR. To ensure network connectivity for all data plane nodes, \emph{DENIS-SDN} employs \emph{Connectivity Detector (CODET)}, a mechanism that ensures that each node within a network slice maintains at least one viable path to transmit data to the border router. 

At the infrastructure layer, an OpenFlow-like protocol enhances communication quality by mitigating radio interference through radio spectrum slicing. This approach involves using alternative radio communication channels for each slice, resulting in improved performance and reduced interference in radio communication. To achieve that, \emph{DENIS-SDN} implements dynamic change of radio channels per network node using the out-of-band control communication channel inherited from \emph{VERO-SDN}. \emph{DENIS-SDN} mechanisms such as \emph{NDC} and \emph{CODET}, along with novel programmable SDN architectural characteristics, provides minimized control overhead, robust operation, and the flexibility to accommodate various UDIoT application requirements, as described in the following sections. 

We showcase the complete \emph{DENIS-SDN} solution and verify the effectiveness of its mechanisms through a comprehensive evaluation utilizing scenarios inspired by real-world use cases. Our evaluation compares logically or physically sliced network operations (i.e., that either conceptually separate the network nodes into slices based on data forwarding rules prioritizing relay nodes from the same slice or by assigning a distinct radio channel to each slice for physical isolation) against traditional non-sliced scenarios for various network topology density and traffic setups. The results highlight the improvements of \emph{DENIS-SDN} in terms of robust packet delivery and improved QoS. 

This paper is organized as follows: in Section~\ref{seq:usecase}, we propose a realistic use-case scenario, emphasizing the impact of our work; Section~\ref{relatedwork} gives an overview of related WSN protocols addressing ultra-dense topology issues;  Section~\ref{seq:framework} details \emph{DENIS-SDN} components, algorithms, and its operation; Section~\ref{experiments} provides an extensive evaluation through simulated scenarios, illustrating the results and achievements of our research proposal; and, finally, Section~\ref{seq:conclusion} concludes the paper, while also discussing our next steps.

\section{Use-case Scenario}
\label{seq:usecase}

\emph{Arenas} or \emph{Indoor Stadiums} are well-known areas that gather a huge crowd to attend sports or other entertainment events. In modern arenas, there are numerous systems and devices that need to be monitored and controlled in real-time to ensure smooth operations, enhance the attendee experience, and optimize resource utilization. 

To further motivate our proposal, we consider a smart crowd management and safety scenario where UDIoT networks can support a large number of connected devices, such as access control systems and occupancy sensors, which can be strategically placed throughout the stadium to monitor crowd movement, detect potential safety hazards, and manage spectator flow during events. This data can be processed in real-time, and insights can be generated to help stadium operators make informed decisions to ensure crowd safety and security.

\begin{figure}[hbt!]
\centering 
\includegraphics[width=3.4in]{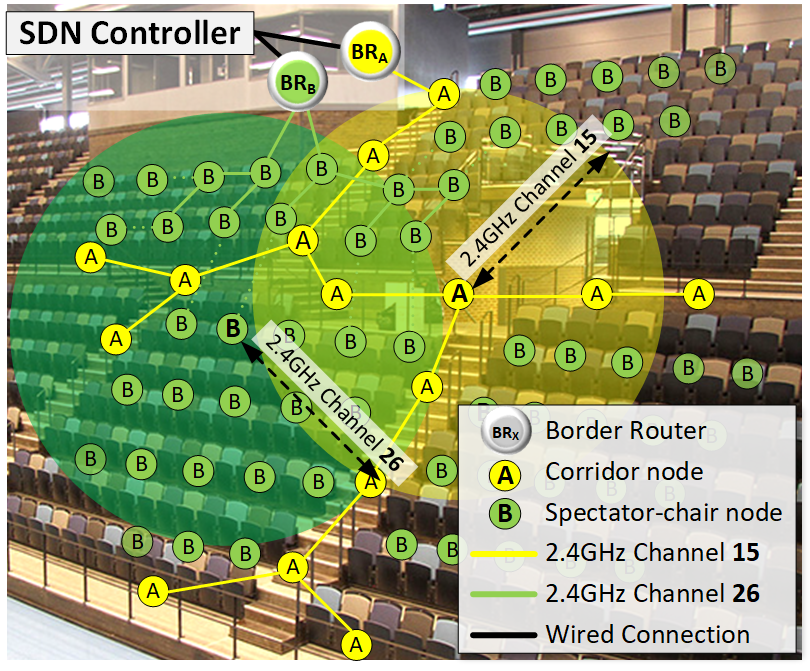}
\caption{\label{fig:ArenaUseCase} Sports Arena use case -- Ultra-Dense topology of sensor nodes detecting occupancy of spectator chairs and arena corridors.}
\end{figure}

In Fig.~\ref{fig:ArenaUseCase}, we depict a scenario where smart spectator-chairs sense their current occupancy status along with other measured data, e.g., environmental or biometric data, and deliver this information to centralized monitoring systems that, in their turn, trigger actuators such as indicator lights or alerting sounds. Moreover, similar occupancy sensor nodes are placed for surveillance along the bleacher's corridors and stairs for public safety, e.g., by detecting people or objects that block their access for a duration of time and activating alarms accordingly. 

This scenario is characterized by a multitude of sensor nodes placed in close proximity to each other. Considering the difference in the significance of the transmitted data from the smart chair sensors versus the corridor surveillance sensors, our proposed networking solution aims to prioritize data delivery based on the role of the sensor. For instance, ensuring reliable data delivery for nodes responsible for spectators' health and safety is paramount. Furthermore, given the high radio interference resulting from the density of network nodes, it is crucial to ensure robust data delivery despite these challenges.

To address the above challenges, \emph{DENIS-SDN} utilizes its \emph{NDC} mechanism to slice the network into two: i) Slice A, the corridor nodes depicted in yellow color in Fig.~\ref{fig:ArenaUseCase}, and ii) Slice B, the smart-chair nodes depicted in green color in Fig.~\ref{fig:ArenaUseCase}. This involves configuring the IEEE 802.15.4 radio transceivers to channel 15 for Slice A, and to channel 26 for Slice B.

The current \emph{DENIS-SDN} version allows the network administrators to configure in real-time slices based on predefined node roles and characteristics. As such, although not presented in this use-case due to space limitations, the administrators could deploy network slices based on odd and even chairs or differentiate between smart chairs for people with special needs and normal chairs to reduce interference and enhance PDR even further. Finally, although the importance of guaranteeing minimum connectivity paths among nodes in each slice may be less critical in UDIoT scenarios due to the plethora of available nodes, we acknowledge that the responsibility currently lies with the administrator. Nevertheless, to support administrators, we introduce \emph{CODET}, an algorithmic mechanism that verifies network connectivity within each network slice by ensuring that every node has at least one path to the border router, thus meeting the minimum connectivity requirements.

In Section~\ref{experiments}, we elaborate on this use case by presenting network configuration details and simulated results for a scenario with 97 nodes and two border routers in various topology deployments. 

\section{Related Works}
\label{relatedwork}
Initial studies in the realm of WSN, such as \cite{ortiz2013fuzzy}, \cite{kim2014intercluster}, and \cite{prabhu2012research}, delved into the implications of dense topologies on data forwarding. These works proposed routing solutions that leverage fuzzy-logic algorithms, intercluster Ant Colony optimization algorithms, and decentralized clustering algorithms to tackle the challenges posed by dense network configurations, respectively.
As the IoT incorporates WSN, it inherits the  difficulties posed by dense network topologies. Today's UDIoT deployments, as exemplified in Section~\ref{seq:usecase}, exacerbate these challenges even further. As such, several recent works propose and investigate potential solutions. 

Sharma~\cite{sharma2018ultraIoT} proposes a novel approach for scheduling sporadic device transmissions in ultra-dense IoT access networks to address data traffic congestion. In this approach, IoT devices adjust transmission timings using an embedded cache to minimize peak data traffic, reducing radio resource demand. 

Guo et al.~\cite{guo2018mobile} investigate mobile edge computing offloading in ultra-dense IoT networks to minimize computation overhead while satisfying wireless channel constraints. They propose an optimal enumeration offloading scheme as a benchmark and a game-theoretic greedy approximation offloading scheme as a practical solution. Likewise, authors in~\cite{guo2019energy} delve into energy-aware task offloading in ultra-dense IoT networks, proposing an iterative searching-based algorithm that jointly optimizes task offloading, frequency scaling, and power allocation using Difference of Convex programming. Both works are validated through numerical results demonstrating the need for computation offloading across multiple edge servers.

Nadif et al. ~\cite{nadif2022traffic} proposes a traffic-aware power allocation algorithm for ultra-dense small cell NB-IoT networks utilizing stochastic geometry analysis and mean-field optimal control. IoT devices cluster around small base stations and adjust transmit power to minimize average utility under fluctuations. The approach estimates network interference with spatial randomness and IoT traffic and the simulation results show constant complexity regardless of density or traffic. 

Authors in~\cite{zhou2021performance} explore a CSMA-type random access scheme with noisy channels for an ultra-dense IoT monitoring system, where multiple devices compete to transmit status packets to receivers over noisy channels. The study considers cases with and without transmission feedback and proposes three policies. A mean-field approximation framework is developed to analyze the system's performance in a large population scenario. Simulation results confirm the accuracy of the derived closed-form expressions for average age of information, even with a few devices.

Moy et al.~\cite{moy2020decentralized} present IoTligent, a decentralized solution to address radio collisions in dense IoT unlicensed bands that aims to extend battery life and mitigate collision issues utilizing machine learning algorithms implemented on IoT devices. The authors suggest employing multi-armed bandit algorithms and demonstrating their effectiveness through a proof-of-concept using LoRa devices in a LoRaWAN network in laboratory conditions. 

In~\cite{emu2021dso} authors advocate for an intelligent Deep Q-Network driven service function chain (SFC) orchestration called DSO aiming to reduce fabrication costs, increase profit margins for providers, and maintain QoS in the context of massive IoT deployments. The DSO model emphasizes sharing already deployed network functions instead of creating new instances, resulting in improved resource utilization. Extensive simulations show that the proposed DSO model significantly reduces running time and achieves near-optimal resource utilization in dense IoT networks.

Although the previously mentioned works address certain challenges in dense and ultra-dense deployments, none consider the timely advancements the SDN paradigm brings in the IoT context. In light of this, we propose \emph{DENIS-SDN}, an SDN platform with mechanisms specifically tailored to address relevant scenarios. To the best of our knowledge, \emph{DENIS-SDN} is the only SDN solution for UDIoT networks. In the following section, we provide a detailed description of the system and its mechanisms.

\section{The DENIS-SDN Framework}
\label{seq:framework}

This section introduces the functional framework and architectural planes of \emph{DENIS-SDN}, followed by an in-depth explanation of the proposed network density management mechanisms used to improve network operation and performance.

\subsection{Architecture Overview}
\label{seq:Architecture}
\begin{figure}[hbt!]
\centering 
\includegraphics[width=3.4in]{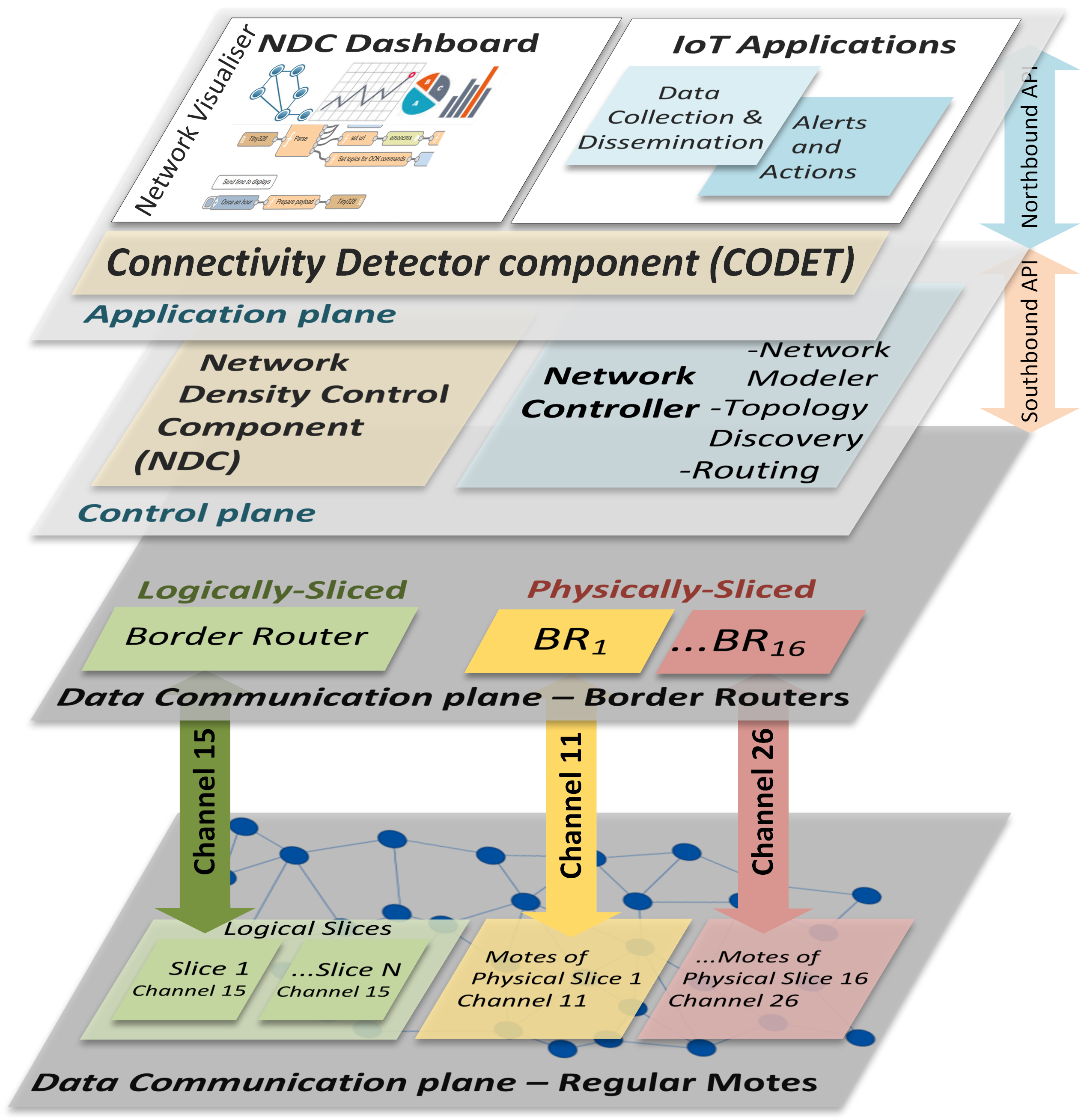}
\caption{\label{fig:Architecture} \emph{DENIS-SDN} architecture}
\end{figure}

Figure \ref{fig:Architecture} presents an overview of the architectural structure of \emph{DENIS-SDN}. The architecture follows the typical three-tier SDN paradigm \cite{ONF2016SNDArchitecture} and consists of the following planes, described from top to bottom:

\begin{enumerate}
\item\emph{Application plane} offers high-level network management and monitoring functionalities, i.e., configuring the number of slices and the nodes in each slice, facilitated by the \emph{Dashboard}, a flexible graphical user interface (GUI). It also provides access to various IoT applications such as Data Collection, Alerts and Actions, and Data Dissemination \cite{gigli2011IoTapps}.  
Furthermore, it incorporates the \emph{CODET} component, a network administration tool that assists in maintaining a high-level QoS in terms of network connectivity between all network nodes and the border router, discussed in detail in Section~\ref{seq:CODET}.

\item\emph{Control plane} includes the \emph{DENIS-SDN Controller}, which manipulates an abstracted anatomy of the infrastructure network. It utilizes sophisticated \emph{Network Control Algorithms} specifically designed for  WSN environments. The control plane maintains an abstract view of the network connectivity, known as the \emph{Network Graph}, through the \emph{Network Modeler} module. The network modeler employs a hybrid topology discovery process adapted to sliced network topologies, employing local (i.e., part or a slice of the network) and global topology discovery algorithms.
It performs network routing and flow control decisions using proactive and reactive flow establishment methods.
It also dynamically adjusts the data-plane protocol parameters and adapts the controller within the dense topology context using the \emph{NDC} component, described in detail in the following subsection. \emph{DENIS-SDN Controller} is implemented in Java and follows a modular, scalable approach. It supports multiple border-router motes and, through its modular architecture, allows the straightforward addition of new algorithms and intelligent functionalities. 

\item\emph{Infrastructure plane} consists of regular and border-router IoT motes. The \emph{DENIS-SDN} infrastructure plane is inherited from the \emph{VERO-SDN} protocol, which is described in detail in our recent research work \cite{theodorou2020VERO-SDN}. This inheritance allows \emph{DENIS-SDN} to leverage the robust and low-overhead network control capabilities of \emph{VERO-SDN} and utilize its dual network stacks for out-of-band radio network control. In detail, \emph{DENIS-SDN} infrastructure plane provides standardized low-power wireless communication and media access control through the IEEE 802.15.4 Physical and Media Access Layers. IEEE 802.15.4 standard~\cite{ieee802} at $2.4$ GHz frequency band defines $16$ radio channels numbered from $11$ to $26$. Each channel is $2$ MHz wide and spaced by $5$ MHz apart to ensure that there is always at least $1$ MHz of separation between any two channels, which is enough to prevent interference. With the $2.4$ GHz band being a crowded spectrum, the $16$ channels provide the network administrators with a wide range of configuration options. 
\emph{DENIS-SDN} capitalizes on this advantage by utilizing the $16$ channels to partition the network into physically isolated slices, ensuring a low-interference radio environment and consequently can utilize a maximum of $16$ physical slices.
The firmware of IoT devices is implemented in the C programming language for Contiki OS v3.0~\cite{dunkels2004contiki}. The choice of Contiki is based on its open-source nature, extensive documentation, widespread use in relevant research initiatives, and the support of both real experimentation and simulations.
\end{enumerate}

The communication between the planes of \emph{DENIS-SDN} occurs via the \emph{Northbound} and \emph{Southbound} APIs, utilizing JSON messages. Detailed technical specifications of these APIs can be found in~\cite{DENIS-SDN2023github}. 

In the following two subsections, we describe each of the \emph{NDC} and \emph{CODET} mechanisms.

\subsection{DENIS-SDN Network Density Control component}
\label{seq:NDC}
The \emph{Network Density Control (NDC)} component augments the SDN controller with management capabilities tailored for high-density networks. It aims to enhance communication quality by partitioning the network into smaller segments. In these segments, data transmission is efficiently managed either at a logical level through optimized data packet routing or at a physical level, where each network segment transmits information on a different communication channel.

\begin{figure}[hbt!]
\centering 
\includegraphics[width=3.5in]{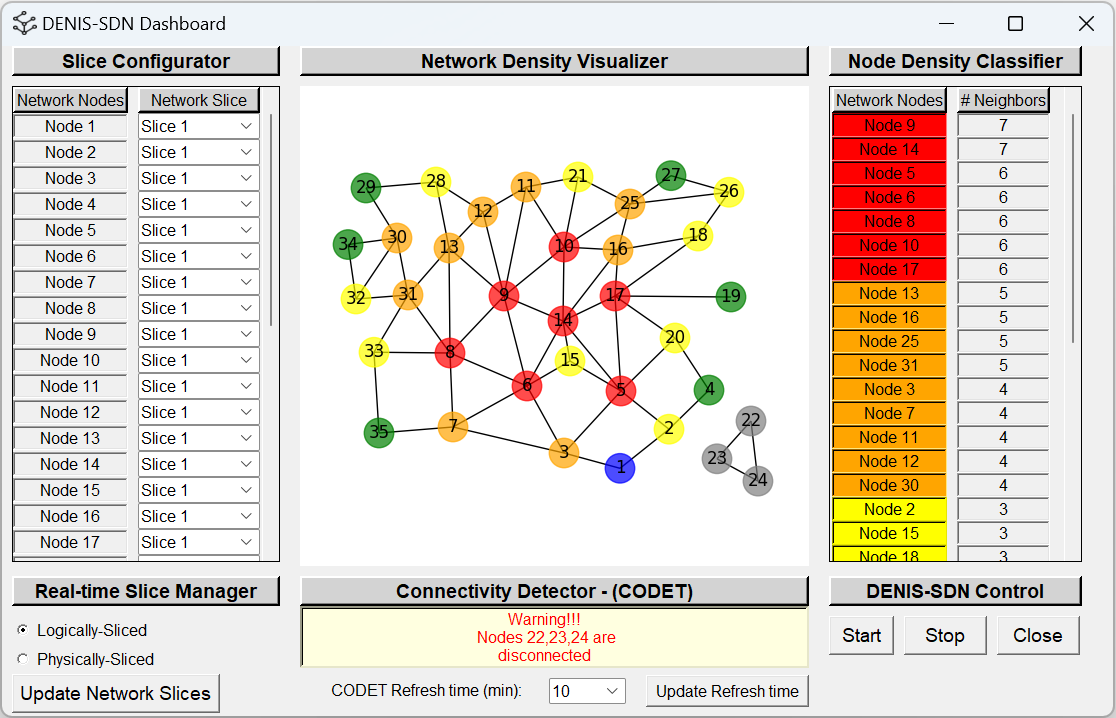}
\caption{\label{fig:Dashboard} \emph{DENIS-SDN} Dashboard}
\end{figure}

In the current version of \emph{DENIS-SDN}, the responsibility for setting up the network slices lies with the network administrator. 
To enhance the administration's decision-making process, \emph{NDC} incorporates the \emph{DENIS-SDN Dashboard}, a user-friendly graphical interface operating at the Application Layer. This dashboard is equipped with a comprehensive range of network monitoring tools and mechanisms, as illustrated in Fig.~\ref{fig:Dashboard}, and described in detail below:
\begin{itemize}
    \item \emph{Slice Configurator}, allows the network administrator to input configuration data for making network slicing decisions. This includes specifying the slice assignment for each node within the network. To enhance usability, nodes that have not been explicitly assigned to any slice by the administrator are automatically included in the default slice, simplifying the configuration process.

    \item \emph{Real-time Slice Manager} allows the administrator to manage network slices in real-time and dynamically make instant adjustments to the slice configuration. To achieve that, \emph{NDC} is utilizing the \emph{DENIS-SDN} sub-GHz out-of-band control channel that can communicate and configure the radio frequency of any of the network nodes in the range of the border router.

    \item \emph{Node Density Classifier} is a real-time mechanism that categorizes network nodes based on their adjacency information using a traffic light RAG rating coloring scheme. It provides valuable insights for network management and optimization, with red nodes indicating high adjacency load, amber and yellow nodes indicating medium load, and green nodes indicating low adjacency load. The classification is percentage-based rather than fixed numbers, allowing adaptability to different density conditions and drawing attention to nodes requiring administrative attention.

    \item \emph{Network Density Visualizer} is a real-time graphical tool that visually represents the network connectivity graph, highlighting regions with high node density. It enables administrators to quickly identify areas of concentrated nodes, aiding in network analysis, capacity planning, and identifying potential congestion points.
\end{itemize}

Yet the main contribution of the \emph{NDC} lies in its ability to effectively coordinate and manage multiple data-plane networks using a single control-plane mechanism. While the \emph{DENIS-SDN Controller} inherits topology discovery and data forwarding mechanisms from \emph{VERO-SDN}, these mechanisms were originally designed for a single set of data-plane nodes. However, through the \emph{NDC} component, the \emph{DENIS-SDN Controller} can orchestrate these mechanisms to operate on multiple networks simultaneously, i.e., network slices. Specifically, \emph{NDC} is responsible for generating and maintaining the \emph{Network Connectivity Graph} $G$ within the \emph{Network Modeler} module, utilizing the topology discovery processes of the \emph{DENIS-SDN Controller}. The \emph{Network Connectivity Graph} depicts a snapshot of network nodes and their corresponding edges in a graph structure that defines radio connectivity among nodes. Within the \emph{Network Modeler}, the \emph{NDC} manages multiple network graphs, with each graph $Gs_i$ corresponding to a specific network slice $s_i$.

Based on the administrator's decision, \emph{NDC} operates in either Logically-Sliced or Physically-Sliced mode, offering distinct functionalities:
\begin{itemize}
    \item In \emph{Logically-Sliced} mode, a single radio channel is utilized for all data-plane nodes, while separate connectivity graphs are maintained for each network slice. The Controller responds to routing requests by making data forwarding decisions that involve relay nodes from the same slice, prioritizing data transmission efficiency.
    \item In \emph{Physically-Sliced} mode, each network slice is assigned a distinct radio channel for its data-plane nodes, allowing for up to 16 network slices as specified by IEEE 802.15.4. The \emph{NDC} component within the Controller handles each slice individually, implementing separate topology control and data forwarding decisions for optimal management. This mode offers the advantage of physical isolation between data-plane nodes of different slices, reducing radio interference and improving overall network performance.
\end{itemize}

\subsection{DENIS-SDN Connectivity Detector component}
\label{seq:CODET}
In multi-hop networks, ensuring that each node has at least one path through peer nodes to transmit its data to a destination node is of utmost importance. The existence of these paths is strongly related to the network topology and the number of nodes used to relay the data. This concern is generally of minor importance in dense and ultra-dense networks due to the abundance of network nodes, enabling alternative paths to be formed. However, in our solution, where the network is segmented into smaller slices based on criteria that may not directly correlate with connectivity, such as specific application needs, like in our use-case scenario, the bleacher's corridor nodes, guaranteeing seamless connectivity becomes a significant concern.
Therefore, we have developed the \emph{Connectivity Detector (CODET)} component, a network administration tool that operates at the Application-layer. Its primary responsibility is to ensure the minimum connectivity requirements are met within each slice by verifying that every node has at least one viable path connecting it to the border router.
We emphasize that the operation of this mechanism is made possible in our solution due to its SDN-based approach and the out-of-band control communication capability. 

In detail, \emph{CODET} is employed by the administrator during the initial phase of network slice configuration to ensure the successful connectivity of each slice. Moreover, \emph{DENIS-SDN} use \emph{CODET} in real-time to perform periodic checks on each network slice to detect potential communication disruptions due to malfunctions. These checks occur every $10\ min$ by default, although the network administrator can customize the interval through the \emph{DENIS-SDN} management GUI (Fig.~\ref{fig:Dashboard}).

\begin{algorithm}
\caption{\emph{CODET} -- Connectivity Detector}
\label{alg:CD}
\SetAlgoLined
\KwIn{$Graph\ G$ -- network-connectivity graph}
\KwIn{$Node\ tN$ -- target node}
\KwOut{$List\ dN$ -- list of disconnected nodes}

\BlankLine
$dN \leftarrow$ []\;
\ForEach{Node $n$ in $G$}{
    \If{$n \neq tN$}{
        $visited \leftarrow$ []\;
        $pathFound \leftarrow$ BFS($n$, $tN$, $visited$)\;
        \If{not $pathFound$}{
            $dN \leftarrow$ $[n]$\;
        }
    }
    \Return $dN$\;
}
\Return $dN$\;

\BlankLine
\SetKwFunction{FBFS}{BFS}
\SetKwProg{Fn}{Function}{: boolean}{end}
\Fn{\FBFS{$start$, $target$, $visited$}}{ \label{alg:line:f1}

    $queue \leftarrow$ [$start$]\;
    \While{$queue$ is not empty}{
        $current \leftarrow$ $queue.pop()$\;
        \If{$current = target$}{
            \Return $true$\;
        }
        \If{$current$ is not in $visited$}{
            $visited \leftarrow$ $[current]$\;
            \ForEach{Node $n$ in $current.neighbors()$}{
                $queue  \leftarrow$ $[n]$\;
            }
        }
    }
    \Return $false$\;
} \label{alg:line:f2}
\BlankLine
\end{algorithm}

In Algorithm~\ref{alg:CD}, we present the operation of \emph{CODET} in pseudocode. The algorithm takes as input the network slice graph $Gs_i$ described in the previous section and the target node $tN$ (e.g., the border router) as inputs. Using a breadth-first search function (lines $\ref{alg:line:f1}$ to $\ref{alg:line:f2}$), it checks if each node in the network slice has at least one path to the target node $tN$. The algorithm returns a list $dN$ that contains the disconnected nodes. If the list is empty, it indicates that the network slice has a full connectivity. If not, an error message notifies the administrator of the connectivity issues.
We highlight that while beyond the scope of this paper, our future work aims to develop an intelligent mechanism capable of resolving connectivity issues by dynamically adjusting the nodes participating in network slices to ensure seamless connectivity.

Finally, to highlight the impact of \emph{CODET}, we note that although the algorithm requires computational power due to its exhaustive  nature, its implementation is carried out within the controller, which benefits from sufficient energy and computational resources. Additionally, the input data required for \emph{CODET}'s operation is information that the controller already maintains for data forwarding tasks. Therefore, \emph{CODET} does not impose additional overhead on network operations.

\section{Evaluation}
\label{experiments}
This section provides our evaluation analysis highlighting the performance advantages of \emph{DENIS-SDN} and its corresponding network slicing control mechanisms. Our goal is to achieve robust packet delivery performance for dense IoT communication environments. Therefore, we carried out a variety of realistic simulations, all considering the discussed use-case in section~\ref{seq:usecase}. The scenario revolves around a data collection application involving $97$ client nodes. These nodes are responsible for gathering measurements from their respective sensors and transmitting them to the control room in the arena for subsequent processing. Considering the application requirements, the client nodes are separated based on their operation into two categories: i) spectator seats and ii) corridor infrastructure monitoring nodes. In our scenario, the node category is considered the primary criterion for slicing the network nodes into two parts. To validate the connectivity of each slice, we utilized the \emph{CODET} mechanism as described earlier. Furthermore, during simulation, \emph{CODET} was employed to detect any potential runtime disconnections that could compromise the integrity of the obtained results.

\begin{table}[hbt!]
\centering
\caption {Network topology density scenarios}
\label{tab:topology-setup}
\begin{tabular}{l c l l} 
    \hline
    \textbf{Density} & \textbf{Distance} & \textbf{Max Neighbors} & \textbf{Deployment Area}\\
    \hline
    {Ultra Dense} & $1\ m$ & $96$ per node & $10 \times 10\ m$\\  
    \hline     
    {Extra Dense} & $2\ m$ & $69$ per node & $20 \times 20\ m$\\   
    \hline       
    {High Dense} & $3\ m$ & $36$ per node & $30 \times 30\ m$\\ 
    \hline       
    {Dense} & $4\ m$ & $20$ per node & $40 \times 40\ m$\\ 
    \hline       
    {Medium Dense} & $4.5\ m$ & $12$ per node & $45 \times 45\ m$\\  
    \hline 
\end{tabular}
\end{table}

In detail, we elaborate on two data traffic scenarios in five different network topology density setups as in Table~\ref{tab:topology-setup}. For each evaluation scenario, we assess \emph{DENIS-SDN} in three operation modes: i) Non-Sliced, ii) Logically-Sliced, and iii) Physically-Sliced, as in Table~\ref{tab:simulation-setup}. 
Due to our simulator limitations, we note that the physically-sliced scenario will require two border router nodes. These nodes need to be configured individually, considering the channel requirements of the respective network slice.
However, we argue that the IoT hardware market offers border routers with multiple antennas simultaneously accommodating different radio channels. 
In addition, we configure the logically-sliced scenario to utilize two border routers to ensure compatibility.

\begin{table}[hbt!]
\centering
\caption {Simulation parameters and configuration setups}
\label{tab:simulation-setup}
\begin{tabular}{l l l} 
    \hline
    \multicolumn{2}{l}{\textbf{Parameters}} & \textbf{Configurations}\\
    \hline
    \multirow{3}{*}{Non-Sliced} 
                    & Border routers & $1$ node\\
                    & Network nodes & $97$ nodes\\
                    & Network topology & Grid\\
     \hline     
     \multirow{5}{*}{Logically-Sliced} 
                    & Border routers & $2$ nodes\\
                    & Slice A nodes  & $21$ nodes\\
                    & Slice A topology  & Linear\\
                    & Slice B nodes  & $76$ nodes\\
                    & Slice B topology  & Grid\\
     \hline   
     \multirow{5}{*}{Physically-Sliced} 
                    & Border routers & $2$ nodes\\
                    & Slice A nodes  & $21$ nodes\\
                    & Slice A topology  & Linear\\                    
                    & Slice B nodes  & $76$ nodes\\
                    & Slice B topology  & Grid\\
     \hline   
    \multirow{1}{*}{Network Deployment}
                    & Density & as in Table~\ref{tab:topology-setup} \\
    \hline    
    \multirow{3}{*}{Network Traffic} 
                    & Data packet size & $128\ Bytes$ \\ 
                    & High & $6$ data-packets/min/node \\ 
                    & Heavy & $10$ data-packets/min/node \\
    \hline
    \multirow{3}{*}{Network Protocols} 
                    & Transport & UDP \\
                    & Network & DENIS-SDN \\
                    & Physical/MAC & IEEE 802.15.4 \\
    \hline
    \multirow{3}{*}{Hardware}
                    & Radio Interface & $2.4\ GHz$ \\  
                    & Radio Range & $25\ m$ \\ 
                    & Mote & Cooja Motes \\
    \hline
    \multirow{2}{*}{Simulation} 
                    & OS / Simulator & Contiki Cooja~\cite{dunkels2004contiki} \\
                    & Duration & $30\ min$ \\
    \hline
\end{tabular}
\end{table}

\begin{figure*}[h!tb]
  \centering
  \subfloat[Network traffic $6$ packets/min per node]
      {\includegraphics[width=0.45\textwidth] {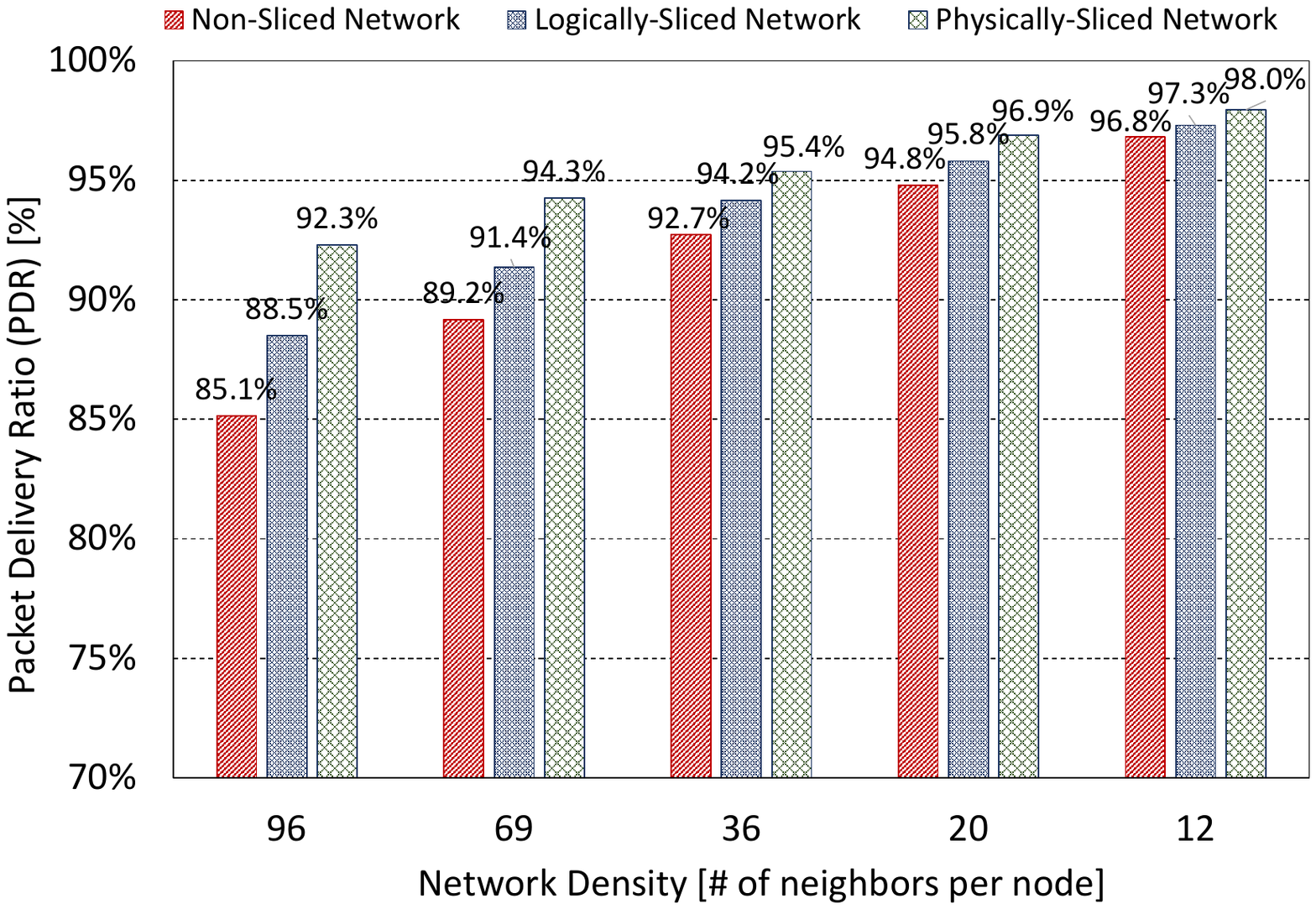}
      \label{fig:6pckt-m-PDR}}
  \subfloat[Network traffic $10$ packets/min per node] 
      {\includegraphics[width=0.45\textwidth] {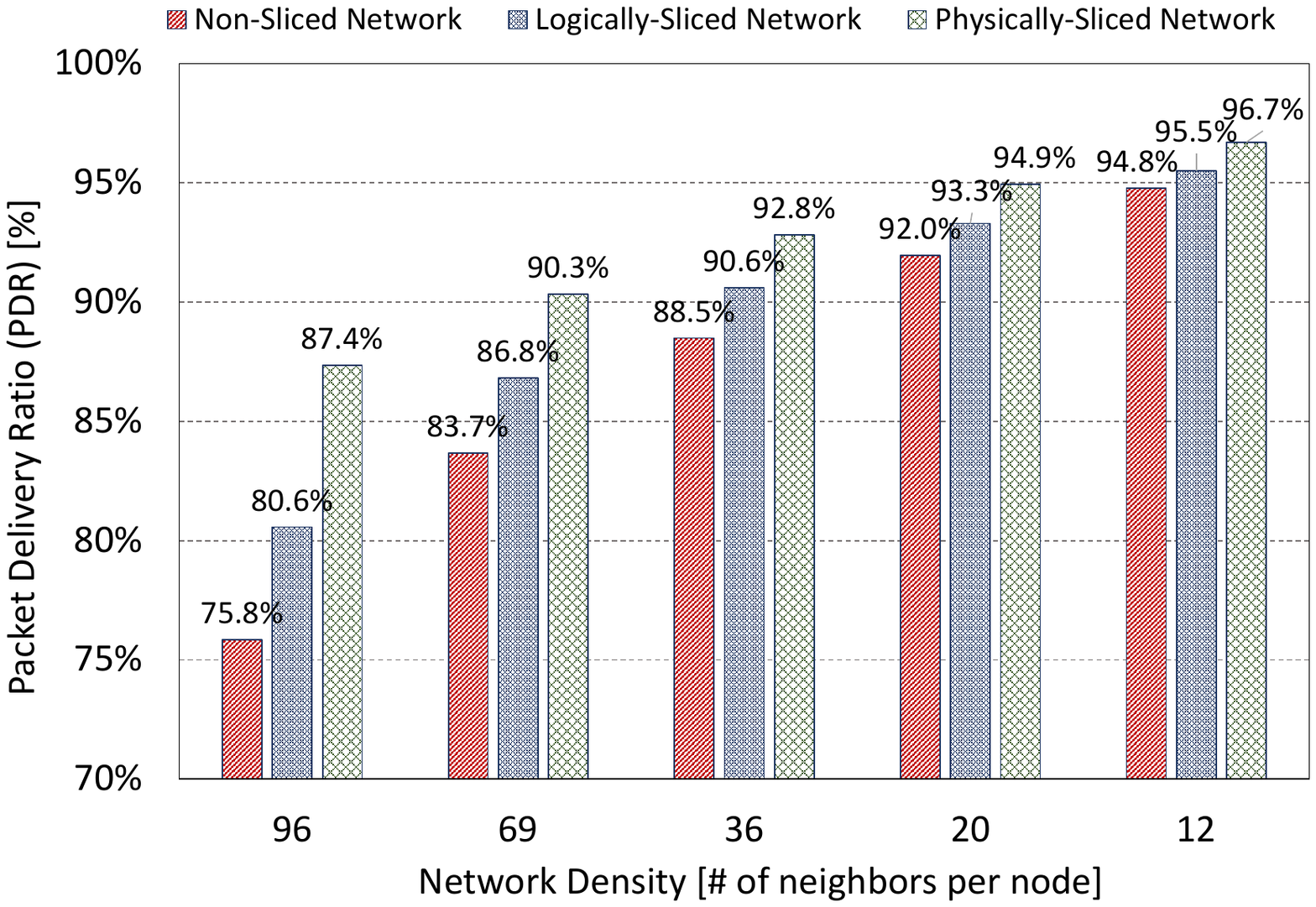}
      \label{fig:10pckt-m-PDR}}
  \hfill
  \caption{Simulation results comparing non-, logically-, and physically-sliced scenarios for \emph{sports arena} use-case.}
  \label{fig:SIM-results}
\end{figure*}

\begin{figure*}
  \centering
  \subfloat[Slice A $6$ pckt/min.]
      {\includegraphics[width=0.45\textwidth] {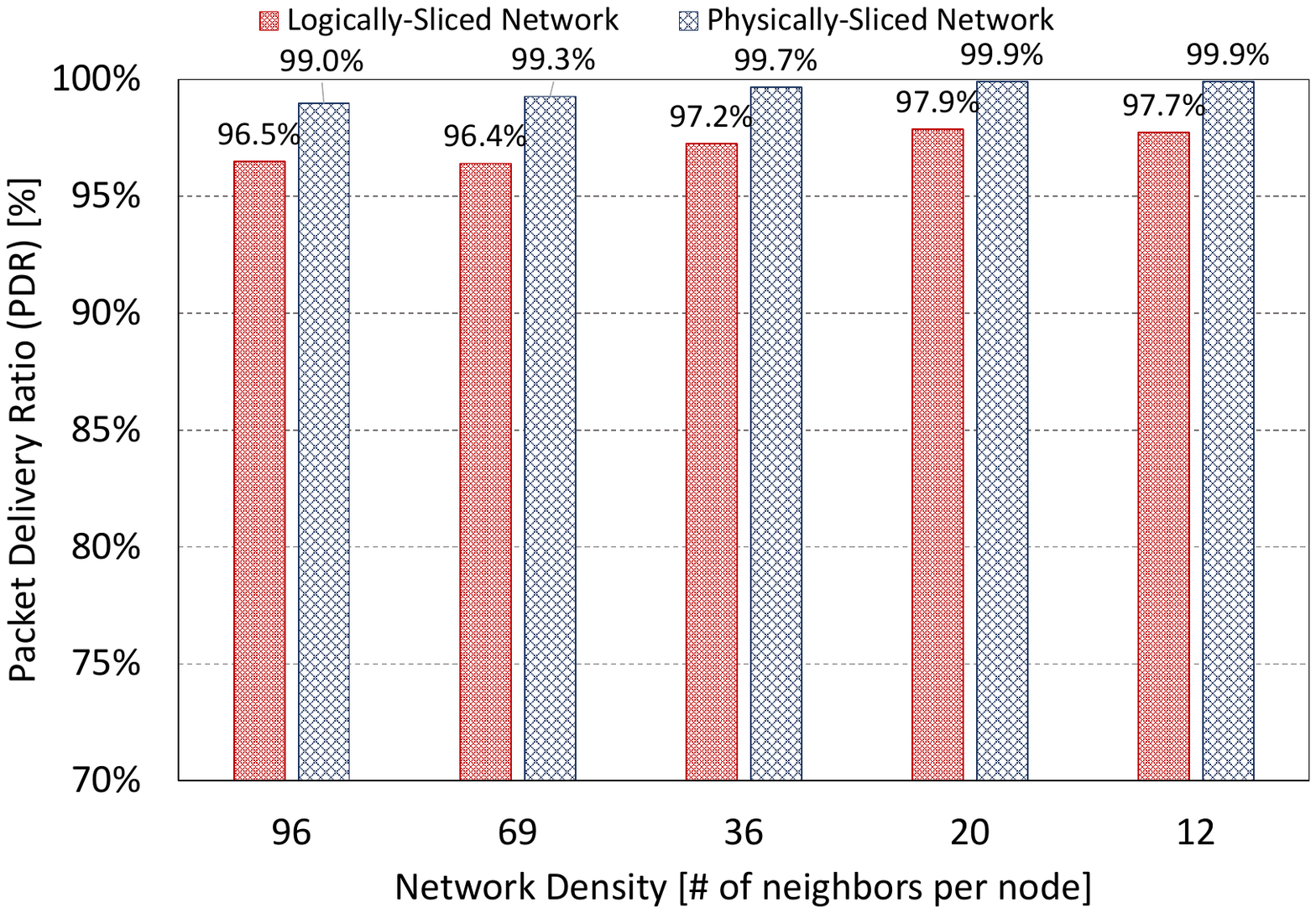}
      \label{fig:A-6-SLICE}}
  \subfloat[Slice B $6$ pckt/min.] 
      {\includegraphics[width=0.45\textwidth] {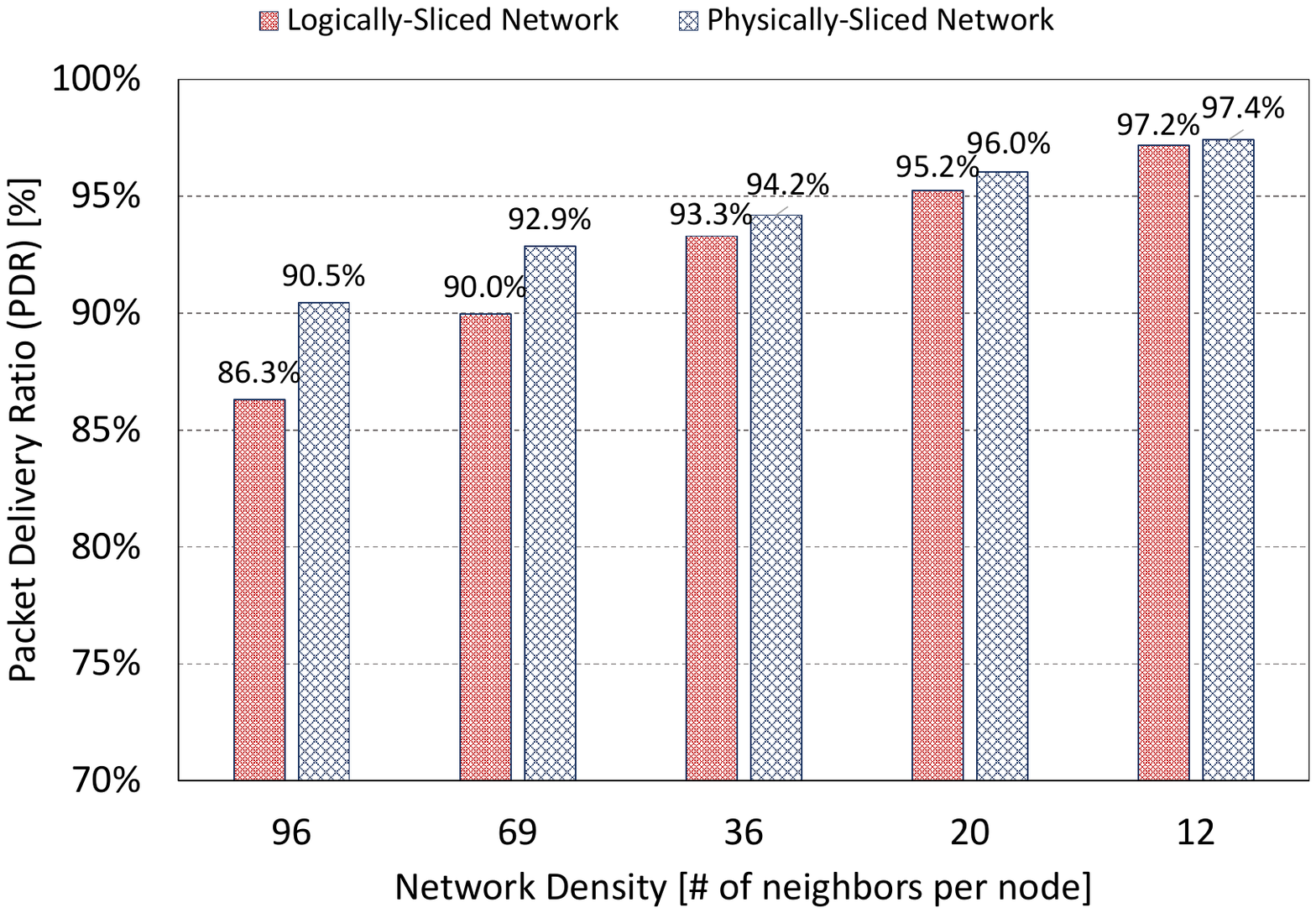}
      \label{fig:B-6-SLICE}}
  \hfill
  \subfloat[Slice A $10$ pckt/min.]
     {\includegraphics[width=0.45\textwidth] {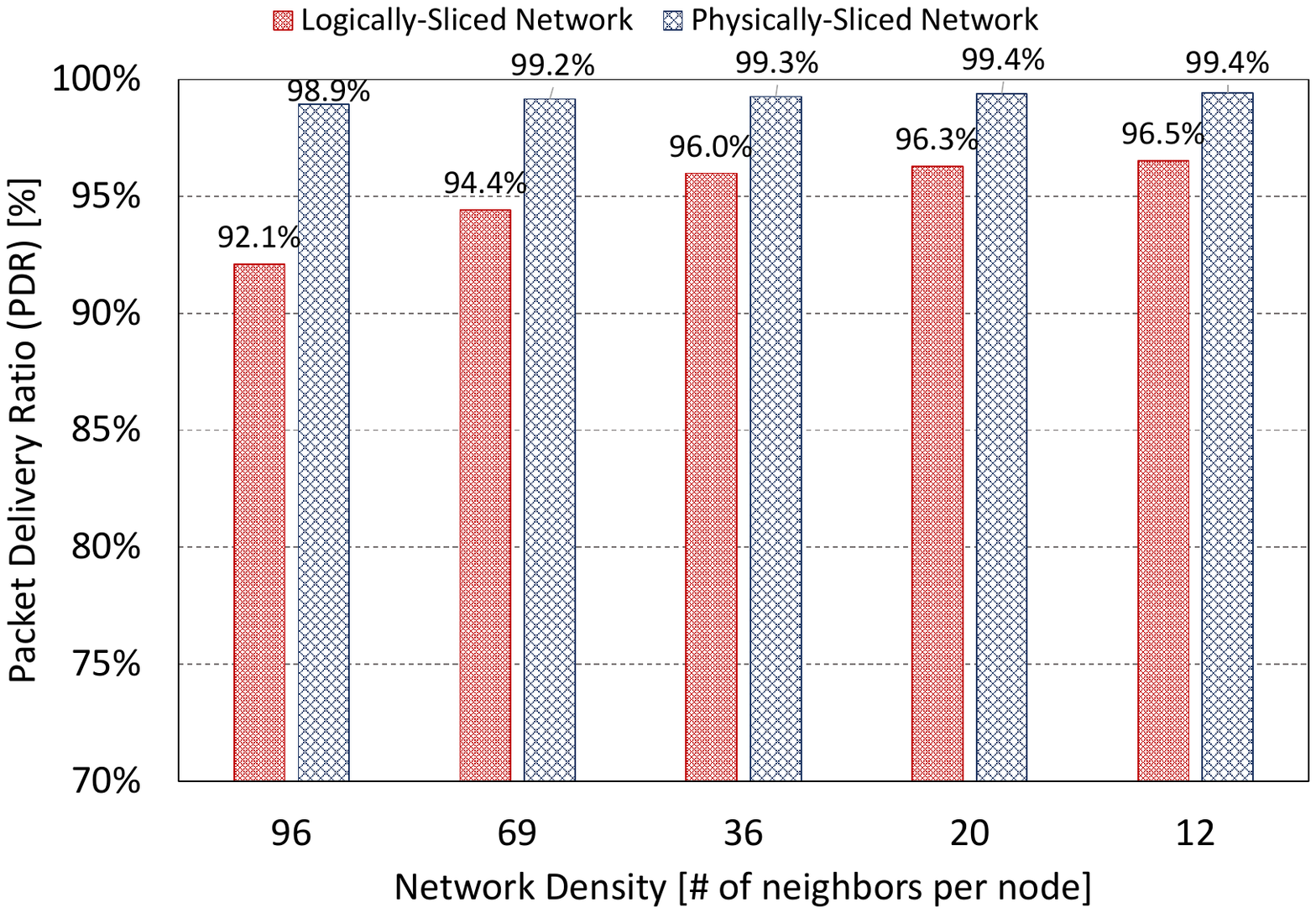}
     \label{fig:A-10-SLICE}}
  \subfloat[Slice B $10$ pckt/min.]
     {\includegraphics[width=0.45\textwidth] {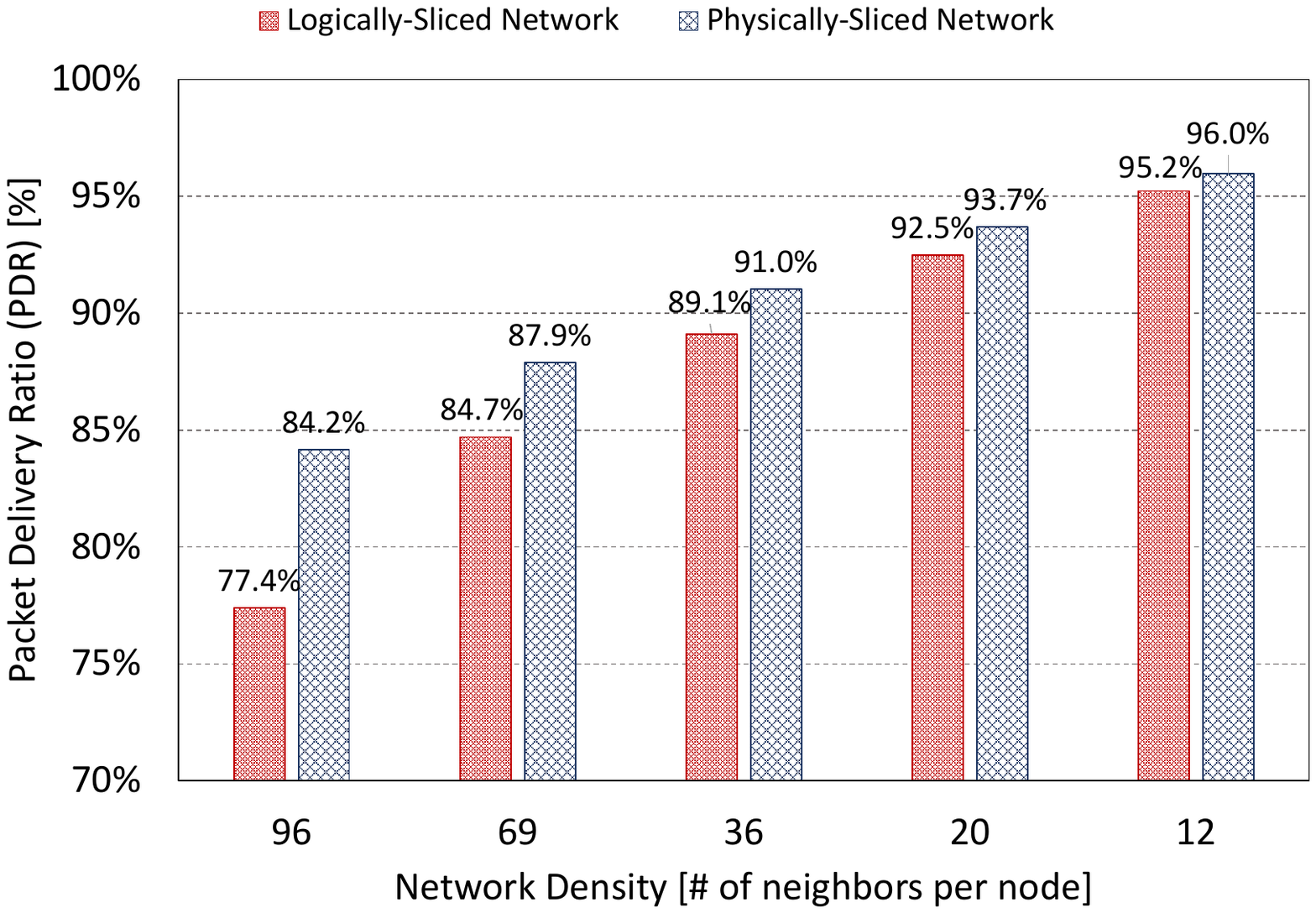}
     \label{fig:B-10-SLICE}}
  \hfill
  \caption{Simulation results of logically- and physically-sliced network per slice for the \emph{sports arena} use-case scenario.}
  \label{fig:MOB-results}
\end{figure*}

In our experiments, we use the \emph{Packet Delivery Ratio (PDR)} metric to measure \emph{DENIS-SDN} reliability in terms of message delivery success ratio. This metric is calculated as the ratio of \emph{received} data messages $R_x$ over \emph{sent} data messages $S_x$, as in (\ref{eq:PDR}). Higher \emph{PDR} values declare reliable transmission, whereas lower values reveal deficiencies in routing processes.  
    \begin{equation} \label{eq:PDR}
        PDR = \frac{\sum R_x}{\sum S_x}
    \end{equation}

In Fig.~\ref{fig:SIM-results}, we illustrate the simulation results for the overall network operation in terms of PDR of two network traffic scenarios, i.e., $6$ and $10$ data packets per minute per node, respectively. In the first traffic scenario (Fig.~\ref{fig:6pckt-m-PDR}), we observe that when the density of the network topology increases, the PDR performance deteriorates. Moreover, the physically-sliced network exhibits the highest PDR across all density scenarios, whereas the non-sliced network yields the lowest. It is worth noting that the drop rate between the non-sliced and physically-sliced operations is double. Specifically, the PDR decreases from $96.8\%$ in the medium-dense topology of the non-sliced operation to $85.1\%$ in the ultra-dense scenario, resulting in an $11.7\%$ decrease, which is twice as much as the $5.7\%$ decrease observed in the physically-sliced network. 
Furthermore, while the logically-sliced network may be less efficient compared to the physically-sliced alternative, it still outperforms the non-sliced operation in every density scenario. 

In the second scenario involving heavier data traffic, i.e., $10\ packets/min\ per\ node$ as depicted in Fig.~\ref{fig:10pckt-m-PDR}, we observe a consistent pattern that aligns with the previously discussed results. Nevertheless, the superiority of the slicing methods becomes even more apparent as the differences in results increase. As such, in the ultra-dense scenario, we observe improved PDR for the logically-sliced operation compared to non-sliced by  $4.8\%$ and for the physically-sliced network by $11.6\%$.

To delve deeper into the evaluation results, our focus shifts to the analysis of PDR per slice, considering both physically-sliced and logically-sliced network operations across all density scenarios. In Fig.~\ref{fig:A-6-SLICE} and Fig.~\ref{fig:A-10-SLICE}, we observe the PDR of Slice-A, which according to the use-case in section~\ref{seq:usecase} is responsible for delivering the data from sensors in critical infrastructure locations, i.e., the surveillance sensors along the bleacher's corridors and stairs of the arena. The results in the physically-sliced operation show $99\%$ and above PDR for all density scenarios and traffic loads. The results justify the suitability of our solution, especially for applications that require reliable transmissions, like in industry and public safety. 

In Fig.~\ref{fig:B-6-SLICE} and Fig.~\ref{fig:B-10-SLICE}, we illustrate the PDR performance of Slice-B nodes, i.e., the intelligent spectator-chair monitoring nodes. We observe that the results for both scenarios, i.e., $6$ and $10$ packets per minute per node, exhibit inferior performance compared to Slice-A for all network density configurations. We claim that the pertinent result is a direct consequence of the number of nodes involved in each slice, i.e., $21$ for Slice-A and $76$ for Slice-B. Therefore, to further optimize the performance of Slice-B, we can consider dividing it into two or more additional slices. As a result, determining the optimal number of slices and nodes per slice holds significant importance. It is worth noting that, at present, \emph{DENIS-SDN} delegates this responsibility to the network administrator while leaving the possibility of automation as a subject for future work.

To summarize, our findings validate the effectiveness of \emph{DENIS-SDN} in dense IoT topologies, enhancing routing efficiency. In our knowledge, \emph{DENIS-SDN} represents the first \emph{SDWSN} solution tailored explicitly for dense IoT environments. We have made \emph{DENIS-SDN} available as an open-source release, anticipating additional enhancements and experimentation from the scientific research community.

\section{Conclusions and Further Work}
\label{seq:conclusion}

In conclusion, the \emph{DENIS-SDN} framework addresses the challenges of dense and ultra-dense network deployments in IoT. \emph{DENIS-SDN} provides efficient and reliable network operation using the SDN paradigm and its innovative \emph{NDC} and \emph{CODET} mechanisms. It offers programmable network slicing, customized routing, and improved resource management, reducing interference and congestion. Through comprehensive evaluations, \emph{DENIS-SDN} has demonstrated robust packet delivery and enhanced QoS, making it a valuable solution for efficient and reliable IoT networks.

Our plans entail conducting extensive investigations to enhance and expand the capabilities of \emph{DENIS-SDN}. As such, we will focus on the following areas:
\begin{itemize}
    \item \emph{Slicing Decision Module}. We intend to augment the functionality of the \emph{DENIS-SDN} Controller by integrating a Slicing Decision Module. This module will utilize an intelligent algorithm to ascertain the most suitable number of slices and network nodes for each slice, taking into account application-specific conditions, desired PDR performance goals, radio connectivity, and load-balancing parameters.
    \item \emph{Extensive experimentation}. We plan to perform large-scale experiments in realistic test beds. Since \emph{DENIS-SDN} already required a significant implementation effort, we left this aspect out as future work.
\end{itemize}

\ifCLASSOPTIONcaptionsoff
  \newpage
\fi

\bibliographystyle{IEEEtran}
\bibliography{citations.bib}

\begin{IEEEbiography}[{\includegraphics[width=1in,height=1.25in, clip,keepaspectratio]{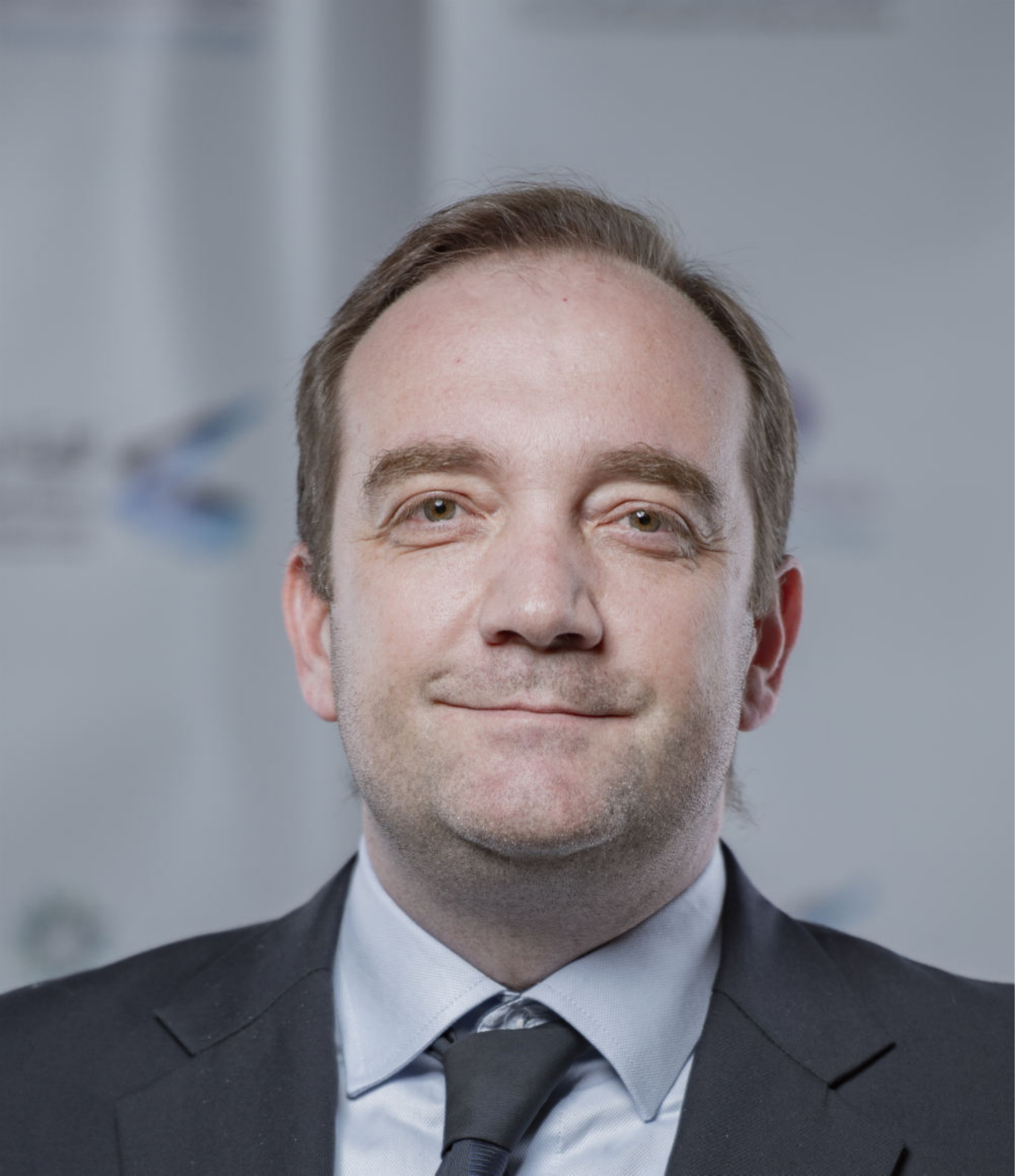}}]{Tryfon Theodorou} (Member, IEEE) holds a Ph.D. degree in Software-Defined Networks for Wireless Devices with Constrained Resources for the Internet of Things from the University of Macedonia, Thessaloniki, Greece. Received an M.Sc. degree in Artificial intelligence knowledge-based systems from the University of Edinburgh, U.K., and an M.Sc. degree in Applied informatics from the University of Macedonia, Thessaloniki, Greece. He has been working in the ICT Sector since 1993. Over the years, he successfully managed and developed various software applications as research or commercial products. He is an active Researcher with several publications. He has participated in multiple international research projects, such as NECOS, WiSHFUL OC2, and MONROE OC2 (H2020). His academic interests include WSNs, SDNs, and IoTNs.
\end{IEEEbiography}

\begin{IEEEbiography}[{\includegraphics[width=25mm,height=32mm, clip,keepaspectratio]{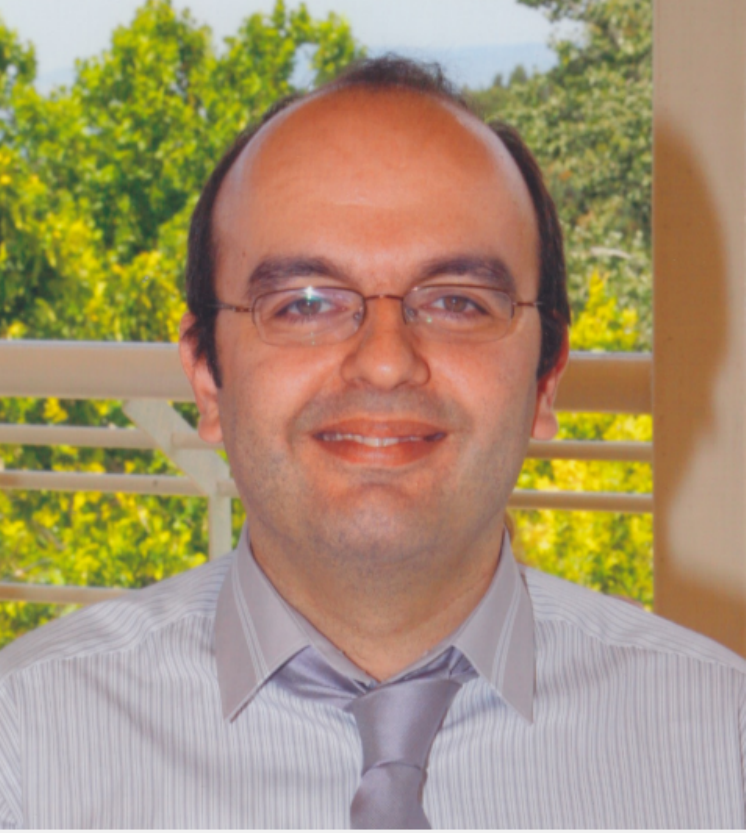}}]{Lefteris Mamatas} (Member, IEEE) is currently an Associate Professor with the Department of Applied Informatics at the University of Macedonia, in Greece. In the past, he has worked as a Researcher at University College London in the United Kingdom, the Democritus University of Thrace in Greece, and DoCoMo Eurolabs in Germany. He has also been involved in numerous international research projects, including CODECO (Horizon Europe), NEPHELE (Horizon Europe), NECOS (H2020), FED4FIRE+ OC4 (H2020), WiSHFUL OC2 (H2020), MONROE OC2 (H2020), and Extending Internet into Space (ESA). He has authored over 80 articles in international journals and conferences, focusing on areas such as Software-Defined Networks, the Internet of Things, and 5G/6G Networks.
\end{IEEEbiography}

\end{document}